\documentclass[twocolumn,aps,tightenlines,floatfix,showpacs]{revtex4}
\usepackage{bm}
\usepackage{graphics}
\usepackage{graphicx}
\usepackage{epsfig}
\usepackage{longtable}
\usepackage{multirow}
\newcommand{\btab}{\begin{tabular}}
\newcommand{\etab}{\end{tabular}}
\usepackage{multirow,amsmath,booktabs}
\begin{document}

\title{Effects of isospin mixing in the $  A=32  $ quintet}

\author{Angelo Signoracci}
\author{B. Alex Brown}

\affiliation{Department of Physics and Astronomy
and National Superconducting
Cyclotron Laboratory,
Michigan State University,
East Lansing, Michigan 48824-1321, USA}

\begin{abstract}
For the $  A=32  $, $  T=2  $ quintet we provide a unified
theoretical description for three related aspects
of isospin mixing:  the necessity of more than three terms
in the isobaric mass multiplet equation,
isospin-forbidden proton decay in $^{32}$Cl,
and a correction to the allowed Fermi $\beta^{ + }$  decay of $^{32}$Ar.
We demonstrate for the first time
that all three effects observed in experiment
can be traced to a common
origin related to isospin mixing of the $  T=2  $
states with $  T=1  $ states.
\end{abstract}


\date{\today}
\maketitle

Recent measurements of the lowest $  T=2  $ states in $^{32}$Cl and 
$^{32}$S have made the
$  A=32  $ multiplet the most precisely measured $  T=2  $ quintet 
\cite{jyfl}, \cite{trim}.
In first-order perturbation theory the masses
in an isobaric multiplet are given by the
isobaric mass multiplet equation (IMME):
$$
M(T_{z}) = a + b T_{z} + c (T_{z})^{2},       \eqno({1})
$$
where $  T_{z} = (N-Z)/2  $ \cite{imme2}.
Multiplets with $  T>1  $ may require
terms of higher order in $  T_{z}  $ that enter in second-order
perturbation theory along with isospin mixing.
The $  A=32  $ multiplet requires a small
but non-zero higher-order term, proportional to  either $  T_{z}^{3}  $ 
\cite{jyfl}
or $  T_{z}^{4}  $ \cite{trim}.

Isospin-forbidden proton decay, another signature
of isospin mixing \cite{ormand}, has been observed from the
$  T=2  $ state in $^{32}$Cl to the low-lying $  T=1/2  $
states in $^{31}$S with a decay width of 20(5) eV \cite{bhat}.

Finally, the superallowed
Fermi $\beta^{ + }$ decay of $^{32}$Ar has been measured recently.
Superallowed 0$^{ + }$ to 0$^{ + }$ Fermi decay
is measured very precisely for many nuclei
and provides the critical data for
extracting the weak mixing angle $  v_{ud}  $
for the KCM matrix \cite{hardy}. The extraction of
$  v_{ud}  $ from the data requires
a small but important correction $\delta_{C}$ due to
isospin mixing.
The correction is smallest for nuclei
near stability, typically 0.5\% or less,
but can be larger for nuclei far
from stability. A correction of
2\% was recently measured \cite{bhat}.

These three isospin-mixing effects, usually studied
and theoretically treated independently, are derived from
the same origin, as evidenced by the $  A=32  $ quintet.
Calculations for energy levels, spectroscopic factors,
gamma decay, isospin-mixing matrix elements, and one-body transition
densities for the multiplet were
obtained in the $  sd  $-shell model space with the USD, USDA, and USDB
interactions \cite{usd}, \cite{bwr}, \cite{usdab}.  The shell model 
code OXBASH \cite{oxw}
was utilized for a full diagonalization of the Hamiltonians.
The USDB interaction will primarily be used for explanation and 
illustration in
this paper, with reference to the other interactions for meaningful 
comparisons.

All calculations were carried out in proton-neutron
formalism. The isospin-mixing interaction is taken from the work
of Ormand and Brown \cite{ormand2}, where in addition to the Coulomb
potential, charge-independence-breaking (CIB) and 
charge-symmetry-breaking (CSB)
interactions were added to the USD Hamiltonian. The CIB strength
was obtained from a one-parameter
fit to the experimental $  c  $ coefficient of the $  T=1  $ IMME
and is consistent with the $  np  $ vs $  pp  $ scattering data 
\cite{ormand}.
The CSB strength was obtained from a one-parameter fit to the
experimental $  b  $ coefficients.

All 0$^{ + }$
states for the $  A=32  $ quintet were calculated. The dominant isospin
of each state was determined by calculating overlaps with
the isospin-conserving part of the interactions.
The lowest $  T=2  $ state was the ground state for $^{32}$Si and 
$^{32}$Ar, the
third 0$^{ + }$ state for $^{32}$P and $^{32}$Cl, and the tenth 
(eleventh,tenth)
0$^{ + }$ state for $^{32}$S using the USDB (USDA,USD) interaction.

\begin{table}
\begin{center}
\caption{Mass excesses of $  T=2  $ states in the $  A=32  $ quintet.}
\begin{tabular}{|r|r|r|r|}
\hline Isobar & T$_{z}$ & $  M_{exp}  $ (keV) & References \\
\hline
\hline
 $^{32}$Si & 2 & -24077.68(30) & \cite{ania} \\
 $^{32}$P  & 1 & -19232.58(12) & \cite{jyfl}, \cite{32p1}, \cite{32p2}, 
\cite{audi}, \cite{endt} \\
 $^{32}$S  & 0 & -13967.57(28) & \cite{trim}, \cite{shi} \\
 $^{32}$Cl & -1 & -8288.34(70) & \cite{jyfl}, \cite{wrede} \\
 $^{32}$Ar & -2 & -2200.2(18)  & \cite{blaum} \\
\hline
\end{tabular}
\end{center}
\end{table}

\begin{figure}
\scalebox{0.5}{\includegraphics{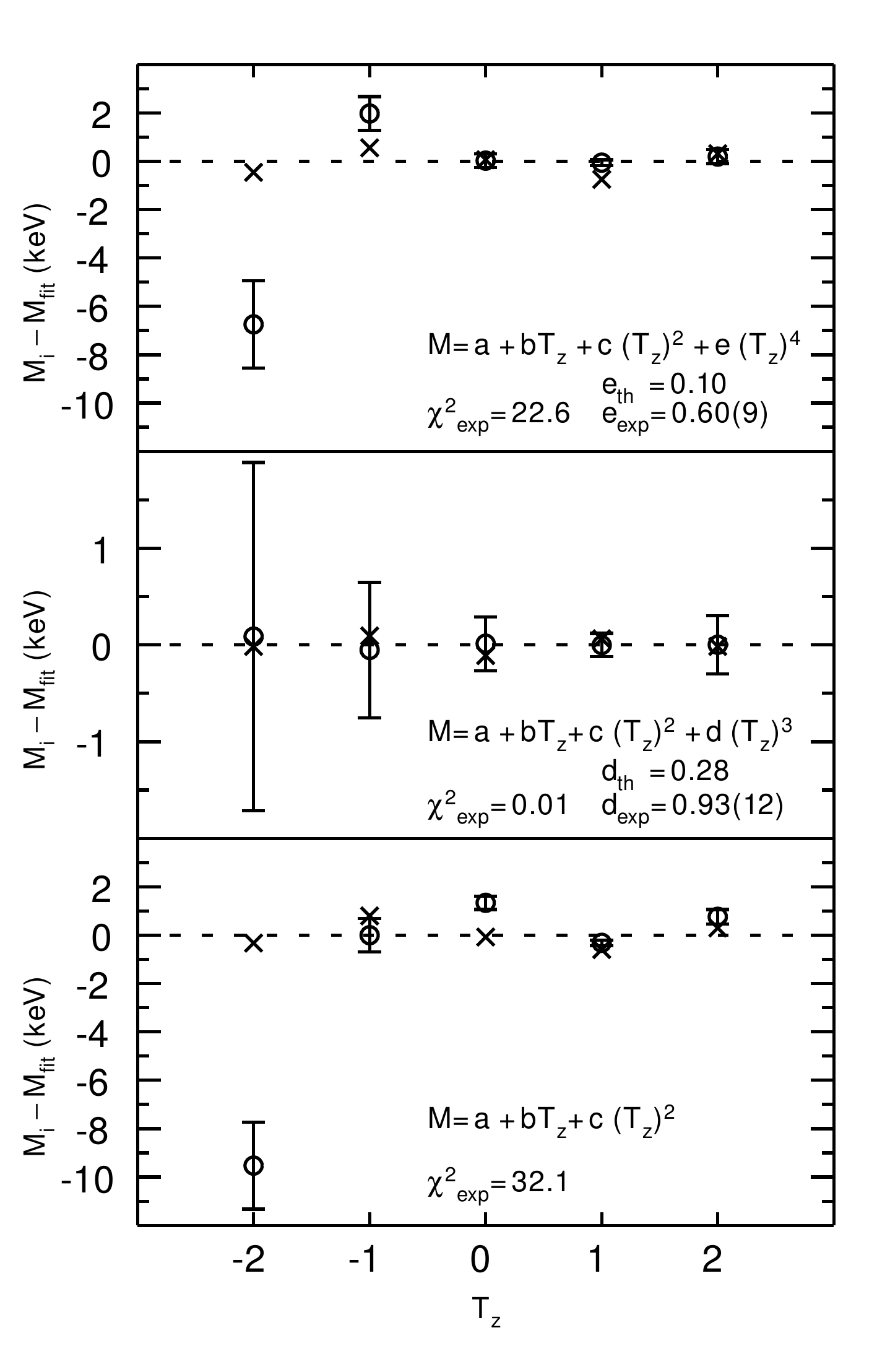}}
\caption{Accuracy of the IMME in the $  A=32  $
quintet with three terms (bottom panel) and with an additional cubic 
(middle)
or quartic (top) term.  Circles correspond to experimental data, with 
error
bars from Table I.  Crosses correspond to theoretical calculations with 
the USDB
interaction.  Note the reduction in scale for the middle panel.}
\label{(1)}
\end{figure}

The experimental masses are given in Table I.  The masses for $^{32}$S 
\cite{trim}, \cite{shi},
$^{32}$Cl \cite{jyfl}, \cite{wrede}, and $^{32}$Ar \cite{blaum} are 
identical to those given in Table I
of \cite{jyfl}.  We have combined the two values for $^{32}$P, 
-19232.46(15) \cite{jyfl}, \cite{32p1}, \cite{32p2} and
-19232.78(20) \cite{audi}, \cite{endt}, into a reduced value based on a 
$\chi^{2}$ fit to a constant.
We use the recent direct measurement at the NSCL for the $^{32}$Si mass
\cite{ania}.  The circles with error bars in the bottom panel of
Fig.\ 1 show the differences in keV between the
experimental masses of the $  T=2  $ states and those
obtained from a fit to Eq.\ (1).  The crosses display the differences 
obtained for the $  T=2  $
energies as calculated with USDB. The differences for both
experiment and theory obtained when $  d \hspace{0.02cm} (e)  $
terms proportional to
 $  T_{z}^{3} \hspace{0.02cm} (T_{z}^{4})  $ are added are shown in the 
middle (top) panel.
For the experimental data, the $\chi^{2}$ value of each fit is given in 
the figure, as well as the best-fit
$  d  $ and $  e  $ parameters.

In Fig.\ 1, it can be seen that a much better fit occurs,
for both USDB and experiment, when a $  d  $ coefficient is used.
We repeated the procedure with $  M_{exp} = -24080.92(5)  $ for 
$^{32}$Si, combining the two
indirect $^{32}$Si masses \cite{paul}, \cite{jyfl} from
Table I of \cite{jyfl} into a reduced value.
While the parameters of the fit and the mass differences change, the 
conclusions are identical.
Again, a $  d  $ coefficient is necessary for a reasonable fit to data, 
producing $  \chi ^{2}_{exp} = 0.58  $ with
$  d_{exp} = 0.53(11)  $.

The most significant difference between theory and experiment in the 
bottom panel of Fig.\ 1 corresponds to
the quality of the fit for $^{32}$Ar ($  T_{z} = -2  $).  A reduction 
in the error bar of $^{32}$Ar,
at least to the level of $^{32}$Cl, would better constrain the fit and 
therefore the $  d  $ parameter.
If we exclude the $  T_{z} = -2  $ point, the $  d  $ term
can be solved algebraically to give $  d_{exp} = 0.95(37)  $, or $  
d_{exp} = 0.41(33)  $ with the
indirect $^{32}$Si mass.  The evidence strongly suggests that the 
three-term IMME does not fit the data
or the USDB calculations.  The uncertainty in the theoretical 
calculations can be assessed by
comparing the results for the three different interactions.  The USD 
interaction result patterns the
USDB behavior but with larger deviations than seen in Fig.\ 1, 
resulting in a greater value of the
necessary coefficient $  d_{th} = 0.39  $.  The USDA interaction
cannot be corrected solely by a $  d  $ coefficient, as large shifts in 
both $^{32}$Cl and $^{32}$S
occur due to isospin mixing.  Both the sign and the magnitude of the 
necessary coefficient give information
about the shifts of the $  T=2  $ states, which can be determined 
theoretically.

\begin{figure}
\scalebox{0.5}{\includegraphics{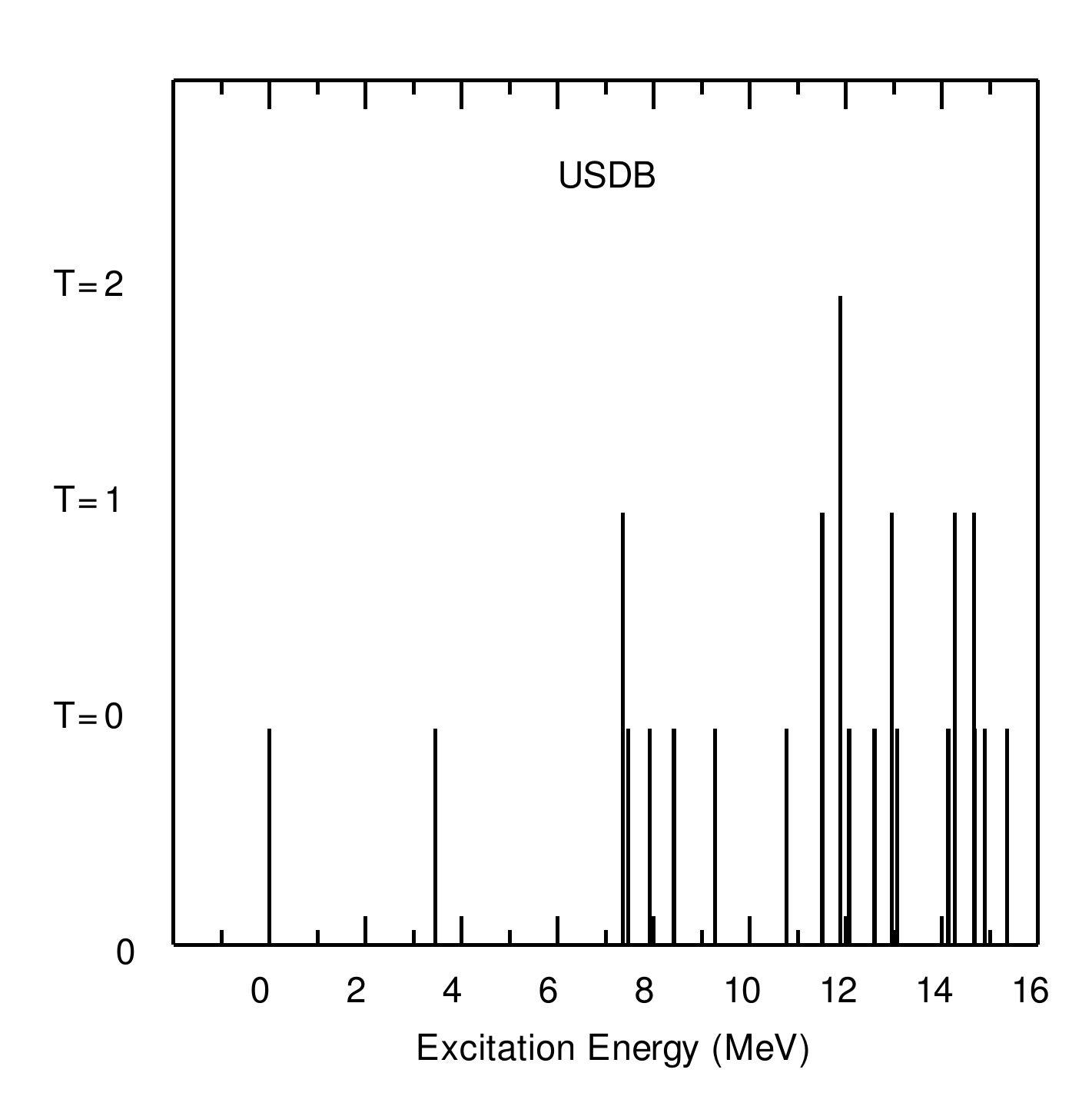}}
\caption{0$^{ + }$ levels in $^{32}$S given by $  T  $ vs. $  E_{x}  $. 
 For $^{32}$P and $^{32}$Cl, the $  T=0  $
levels can be ignored to give an approximate distribution of the 0$^{ + 
}$ states.}
\label{(2)}
\end{figure}

Fig.\ 2 shows the first twenty 0$^{ + }$ levels in $^{32}$S with the 
USDB interaction,
categorized by their values of $  T  $.  The sum of the ground state 
energy
and the excitation energy of the $  T=2  $ state gives the energy to be 
used
in the fit for $^{32}$S.
As seen from the graph, the $  T=2  $ state has nearby states that 
repel it,
shifting its energy.  The shift of the $  T=2  $ state, labeled by 
``a", is approximately given by
$$
\Delta E_{a} = - \sum_{{}i \ne a} \frac{\mid <i\mid V\mid a>\mid 
^{2}}{E_{i}-E_{a}},       \eqno({2})
$$
where $  V  $ denotes the interaction that causes isospin mixing.  $  
E_{x}  $, the
excitation energy of the state, is given by $  E_{a} + \Delta E_{a}  $.
The closest states generally have the greatest effect, and therefore 
most
terms in the sum can be ignored.  The $  T=2  $ state
in $^{32}$S can be shifted by both $  T=0  $ and $  T=1  $ states, 
while the states
in $^{32}$P and $^{32}$Cl can only be shifted by $  T=1  $ states ($  
T=0  $ levels do
not exist in these nuclei).  Table II
shows the significant contributing levels to energy shifts in the $  
A=32  $
quintet for all three interactions.

\begin{table}
\begin{center}
\caption{Energy shifts in the $  A=32  $ quintet for the three 
interactions,
where $  \Delta E_{i}, d_{i}  $, and $  e_{i}  $ are the contributions 
to the energy shift, $  d  $, and  $  e  $
coefficients, respectively, from state $  i  $ (i.e., $  \Delta E_{a} = 
\sum_{i} \Delta E_{i}  $).
The algebraic solution to the five term IMME gives exact values of $  d 
 $
and $  e  $ for each interaction.}
\begin{tabular}{|c|r|r|r|r|r|r|r|}
\hline Interaction & Isobar & T & $\mid<$i$\mid$V$\mid$n$>\mid$ & $  
E_{i}-E_{a}  $ & $  \Delta E_{i}  $ & $  d_{i}  $ & $  e_{i}  $ \\
 & & & (keV) & (keV) & (keV) & (keV) & (keV) \\
\hline
\hline
\multirow{5}{*}{USDB} & $^{32}$P & 1 & 13.14 & 1020 & -0.163 & 0.027 & 
0.027 \\
 & $^{32}$P & 1 & 22.48 & 2387 & -0.215 & 0.036 & 0.036 \\
 & $^{32}$Cl & 1 & 20.92 & -439 & 1.084 & 0.181 & -0.181 \\
 & $^{32}$S & 1 & 17.26 & -377 & 0.633 & 0 & 0.158 \\
 & $^{32}$S & 0 & 7.03 & 182 & -0.294 & 0 & -0.073 \\
\hline
 {\bf Exact} & & & & & & {\bf 0.28} & {\bf -0.07} \\
\hline
\multirow{5}{*}{USDA} & $^{32}$P & 1 & 12.73 & 1025 & -0.154 & 0.026 & 
0.026 \\
 & $^{32}$P & 1 & 20.39 & 2643 & -0.159 & 0.027 & 0.027 \\
 & $^{32}$Cl & 1 & 20.54 & -352 & 1.324 & 0.221 & -0.221\\
 & $^{32}$S & 1 & 12.26 & -302 & 0.405 & 0 & 0.101 \\
 & $^{32}$S & 0 & 19.32 &  -33 & 4.939 & 0 & 1.235\\
\hline
 {\bf Exact} & & & & & & {\bf 0.30} & {\bf 1.40} \\
\hline
\multirow{6}{*}{USD} & $^{32}$P & 1 & 13.32 & 969 & -0.174 & 0.029 & 
0.029\\
 & $^{32}$P & 1 & 19.97 & 1544 & -0.262 & 0.044 & 0.044 \\
 & $^{32}$Cl & 1 & 19.10 & -262 & 1.606 & 0.268 & -0.268 \\
 & $^{32}$S & 1 & 18.18 & -195 & 1.082 & 0 & 0.271\\
 & $^{32}$S & 0 & 13.42 & 475 & -0.399 & 0 & -0.100 \\
\hline
 {\bf Exact} & & & & & & {\bf 0.39} & {\bf 0.03} \\
\hline
\end{tabular}
\end{center}
\end{table}

The excitation energy of the $  T=2  $ 0$^{ + }$ state in $^{32}$S is 
11.885 (11.867,12.011) MeV with the
isospin-nonconserving USDB (USDA,USD) interaction, in reasonable 
agreement with the experimental
value of 12.048 MeV.  However, the nearest $  T=0  $ 0$^{ + }$ state is 
182 (252) keV higher in energy
for USDB (USD) but 33 keV lower in energy for USDA.  There are no known 
experimental 0$^{ + }$ states
above the $  T=2  $ state at 12.048 MeV; the nearest known experimental 
0$^{ + }$ level
is a $  T=0  $ state 118 keV below the $  T=2  $ state.  There is
also an experimental 0$^{ + }$ state 180 keV below without an assigned 
$  T  $ value
which could correspond to the $  T=1  $ state seen 377 (302,195) keV 
below for USDB (USDA,USD).
The proximity of these two states as
calculated via the different empirical interactions has no deep 
underlying cause, but is rather
an incidental effect due to the configurations of the states.  Because 
the shift varies inversely
with the energy, the proximity of the mixing state determines the size 
of the shift
and the observed deviation from the three-term IMME.  This energy 
difference (in
conjunction with the size of the isospin-mixing matrix element)
determines the size of the necessary $  d  $ or $  e  $ coefficient.
In an algebraic solution of an isobaric quintet to the five term IMME
(including both $  d (T_{z})^{3}  $ and $  e (T_{z})^{4}  $ terms), the 
USDB and USD calculations, as well as
the experimental data, result in a small ( $   \leq 0.1   $ keV) $  e  
$ coefficient.  The USDA
calculations require an $  e  $ coefficient of 1.4 keV, even larger 
than the necessary $  d  $ coefficient.
Since the small energy difference in the USDA level scheme results in 
strong mixing in $^{32}$S,
in opposition to the experimental data, we will favor the USDB result.

The decay from the $  T=2  $ state in $^{32}$Cl occurs via two 
processes:
$  \gamma  $ decay and proton emission.  The primary channel of $  
\gamma  $
decay is an M1 transition to the first excited 1$^{ + }$ state in 
$^{32}$Cl
with a branching ratio of 94\%, using the USDB interaction and free 
gyromagnetic
factors.  The calculated width $  \Gamma_{\gamma}  $ is 1.11 eV, in
comparison to the experimental value of $  \Gamma_{\gamma}  $ = 1.8(5)
eV.  The isospin-forbidden proton transition decays to $^{31}$S.  The
reaction has $  Q = 3.45  $ MeV for both USDB and experiment.  Since 
the transition is from a
$  J^{\pi}=0^{+}  $ level, and the proton has $  j = 
1/2^{+},3/2^{+},5/2^{+}  $ in the $  sd  $
shell, decay can only occur to levels in $^{31}$S with those values of 
$  J  $.
Five such levels have $  E_{x} \leq Q   $, all with $  T=1/2  $ using 
the isospin-conserving part of
the Hamiltonians.  With the inclusion of the CIB and CSB interactions, 
isospin mixing occurs
in both the parent $^{32}$Cl and daughter $^{31}$S nuclei.  The small 
spectroscopic factors for
the isospin-forbidden transitions are shown in Table III.
If isospin mixing is only included for $^{32}$Cl, the spectroscopic 
factors are larger.  It is
therefore important to include mixing in both nuclei to account for
interference effects in the wavefunctions.  The decay widths to states 
in $^{31}$S were determined by
$$
\Gamma_{p} = \sum_{j} C^{2}S \thinspace \thinspace \Gamma_{sp},       
\eqno({3})
$$
where $  C^{2}S  $ are spectroscopic factors and $  \Gamma_{sp}  $ is 
the single
particle width of the resonance peak in the reaction $^{31}$S $+$ p
$  \rightarrow  $ $^{32}$Cl.  The single particle widths were 
calculated from
scattering phase shifts in a Woods-Saxon potential \cite{ws}, 
\cite{cham}
with the potential depths chosen to reproduce the resonance energies.  
The
results for the five levels are shown in Table III, but only the $  
1/2^{+}  $ ground state and
first excited state (3/2$^{ + }$) contribute to the decay.
Therefore, $  \Gamma_{p}  $ = 41.4 eV, in comparison to the
experimental value of 20(5) eV from \cite{bhat}.  There is a large
uncertainty in the $  1/2^{+}  $ level, the most important 
contribution, for
two reasons: (i) the calculation of $  \Gamma_{sp}  $ was determined by 
doubling the
width at half max on the low-energy side of the resonant peak, due to 
the large tail in the
resonant reaction at high energy and (ii) the spectroscopic factor 
changes by a
factor of four depending on the interaction used, with $  C^{2}S  $ 
values of 0.00003 (0.00008,0.00012)
for the USDB (USDA,USD) interactions.
Using the same single particle widths for the USDA interaction gives  $ 
 \Gamma_{p} = 92.2  $ eV.
The isospin-mixing matrix element between the $  T=2  $ state in 
$^{32}$Cl and the $  T=1  $ state
below it is approximately the same for all three interactions, as seen 
in Table II.  The
energy change difference varies from 262 to 439 keV, however,
where the range is on the order of the 150 keV global rms energy 
deviation for
the three interactions.  However, a difference of 100 keV in the energy 
denominator
can significantly affect the
width of proton decay due to the amount of mixing.  The USDB energy
difference is greatest and has the smallest proton decay width, in 
better agreement with
experiment.

\begin{table}
\begin{center}
\caption{Widths of the isospin-forbidden proton decay for $^{32}$Cl 
using the USDB interaction.}
\begin{tabular}{|r|r|r|r|r|}
\hline $  J^{\pi}  $ & $  E_{x}  $ & $  C^{2}S  $ & $  \Gamma_{sp}  $ 
(keV) & $  \Gamma  $ (eV) \\
\hline
\hline
 1/2$^{ + }$ & 0.00 & 0.00003 & 1002 & 33.82 \\
 3/2$^{ + }$ & 1.19 & 0.00041 & 18.4 & 7.54 \\
 5/2$^{ + }$ & 2.30 & 0.00001 & 0.3 & 0.00 \\
 1/2$^{ + }$ & 3.20 & 0.00005 & $  \approx  $ 10$^{-6}$ & 0.00\\
 5/2$^{ + }$ & 3.32 & 0.00022 & $  \approx  $ 10$^{-8}$ & 0.00\\
\hline
\end{tabular}
\end{center}
\end{table}

In the $  \beta  ^{ + }$ decay from the ground state of $^{32}$Ar to 
the $  T=2  $,
$  J^{\pi}=0^{+}  $ state in $^{32}$Cl (both members of the $  T=2  $ 
multiplet), the
$  ft  $ value differs slightly from the expected value due to the
isospin mixing in $^{32}$Cl.  The only significant contribution comes 
from the
T=1, $  J^{\pi}=0^{+}  $ state below, the same state influential in the 
isospin-forbidden
proton decay to $^{31}$S and in the deviation from the three-term IMME.
The calculated value of $  \delta_{C}  $ is 0.27\%
(0.40\%,0.63\%) for the USDB (USDA, USD) interactions, allowing us to 
quote a
theoretical value of 0.43(20)\% from an average of the three 
calculations.
Again, the energy difference results in a range of results for an 
isospin-mixing effect.\
From \cite{hardy}, $\delta_{C}$ should be a sum of a charge-dependent 
mixing contribution
(calculated here) and a radial overlap component (1.4\% from 
\cite{bhat}).
The sum of these two
contributions, or 1.8\%, agrees with the experimental values of 
2.1(8)\% \cite{bhat}
and, based on a new mass of $^{32}$Cl, 1.8(8)\% \cite{wrede}.
In all three calculations, nearly the entire remaining strength
feeds to the 0$^{ + }_{2}$ state in $^{32}$Cl, the $  T=1  $ state 
shown in Table II.  The
transition to this state from the ground state of $^{32}$Ar might be
accessible experimentally.

The experimental data for the $  A=32  $ multiplet differs from the 
IMME fit
with three terms, requiring another term for an adequate fit.  Using 
the new direct measurement
of $^{32}$Si \cite{ania}, a $  d  $ coefficient of 0.93(12) is 
necessary.  With the indirect mass of
$^{32}$Si \cite{paul}, \cite{jyfl}, the necessary $  d  $ coefficient 
is 0.53(11).  The
three term IMME similarly does not reproduce the behavior of the masses 
using three different $  sd  $
interactions.  The calculated $  d  $ coefficient is 0.28 (0.30,0.39) 
for the USDB (USDA,USD) interactions.
The USDA calculations also result in a large $  e  $ coefficient due to 
the proximity
of a $  T=0  $ state to the $  T=2  $ state in $^{32}$S.  There is an 
inherent uncertainty
in our method regarding the shift due to isospin mixing on account of 
the global rms deviation
of 150 keV of empirical interactions.  We gain information from using
multiple interactions, but rely on experiment to constrain our choice 
of interaction for comparison.
With the USDB interaction, the decay of the $  T_{z} = -1  $ state in 
the multiplet
occurs primarily by proton emission with $  \Gamma_{p}  $ = 41.4 eV, 
but the gamma
decay with $  \Gamma_{\gamma}  $ = 1.11 eV cannot be neglected.  The 
proton decay width is
approximately double the experimental value, while the gamma decay 
width is in relatively good
agreement with experiment.  The theoretical $  \Gamma_{p}  $ result 
varies significantly with the
interaction used, suggesting a large uncertainty in the calculated 
value.
Regardless of the interaction chosen, the mixing of the $  T=1  $ and $ 
 T=2  $ states
in $^{32}$Cl causes a nonzero isospin-forbidden proton decay to $  
T=1/2  $ states in
$^{31}$S.  The mixing of these same states also accounts for the 
deviation
in the $  ft  $ value of the $  \beta  ^{ + }$ decay of the ground 
state of $^{32}$Ar.
The isospin-breaking correction $  \delta_{c}  $ = 1.8\%
agrees with the experimental value.

These observed aspects of isospin mixing occur in relation to the 
proximity of the levels,
separate from the correlation between the configurations of the states. 
 In the event
of small energy differences and non-negligible isospin-mixing matrix 
elements,
the effects described above will be seen.  The commonness of the 
fulfillment of
these two requirements in other multiplets, such that effects of
isospin mixing occur, cannot be
determined without more accurate theoretical energies or
more complete experimental level schemes.

\vspace{ 12pt}
\noindent 
  {\bf Acknowledgments}
Support for this work was provided from
National Science Foundation Grant PHY-0758099, from the Department of 
Energy
Stewardship Science Graduate Fellowship program through grant number
DE-FC52-08NA28752, and from the DOE UNEDF-SciDAC grant 
DE-FC02-09ER41585.

\end{document}